\documentclass[12pt,preprint]{aastex}

\newcommand{\nc}{\newcommand}

\nc{\M}{\rm { M } }
\nc{\Msun}{\rm { M }_{\sun}}
\nc{\R}{\rm { r }}
\nc{\rv}{{ R_v }}
\nc{\impc}{\rm { h/Mpc }}
\nc{\mpc}{\rm { Mpc/h }}
\nc{\etal}{{\it et al.\ }}
\nc{\xiav}{\bar{\xi}}
\nc{\omav}{\bar{\omega}}
\nc{\bn}{\bar{N}}
\nc{\tP}{\tilde{P}}
\nc{\tF}{\tilde{F}}
\nc{\avg}[1]{\langle{#1}\rangle}
\nc{\abs}[1]{\mid{#1}\mid}
\nc{\T}[1]{\langle{#1}\rangle_C}

\nc{\eg}{{\it e.g.,\ }}
\nc{\ie}{{\it i.e.\ }}

\nc{\bea}[1]{\begin{eqnarray} \mbox{$\label{#1}$}}
\nc{\Section}[2]{\section{#2}\label{#1}}
\nc{\Bibitem}[1]{\bibitem{#1}}
\nc{\Label}[1]{\label{#1}}

\nc{\vev}[1]{\langle #1 \rangle}

\nc{\be}{\begin{equation}}
\nc{\eea}{\end{eqnarray}}
\nc{\ee}{\end{equation}}
\nc{\eeq}{\end{equation}}

\def\lcdm{$\Lambda$CDM~}
\def\ltsima{$\; \buildrel < \over \sim \;$}
\def\gtsima{$\; \buildrel > \over \sim \;$}
\def\simlt{\lower.5ex\hbox{\ltsima}}
\def\simgt{\lower.5ex\hbox{\gtsima}}

\begin{document}
\title{Cosmological Three-Point Function: \\
Testing The Halo Model Against Simulations}

\author{Pablo Fosalba\altaffilmark{1, 2}, Jun Pan\altaffilmark{1,3}, Istv\'an Szapudi\altaffilmark{1}}

\altaffiltext{1}{Institute for Astronomy, University of Hawaii,
2680 Woodlawn Dr, Honolulu, HI 96822, USA} 
\altaffiltext{2}{Instituto de Ciencias del Espacio (IEEC/CSIC), Facultat de Ciencies UAB,
Torre C5 -par- 2a planta,  08193 Bellaterra (Cerdanyola), SPAIN} 
\altaffiltext{3}{School of Physics and Astronomy, University of Nottingham, Nottingham NG7
2RD, UK}

\begin{abstract}

We perform detailed comparison of the semi-analytic halo model 
predictions with measurements in numerical simulations 
of the two and three point 
correlation functions (3PCF), as well as power spectrum and bispectrum.
We discuss the accuracy and self-consistency 
of the halo model description of gravitational
clustering in the non-linear regime and constrain halo model parameters. 
We exploit the recently proposed multipole expansion of 
three point statistics that expresses rotation invariance in 
the most natural way. This not only
offers technical advantages by reducing the integrals required 
for the halo model predictions, but amounts to a convenient way of compressing
the information contained in the 3PCF.
We find that, with an appropriate choice of the halo boundary and 
mass function cut-off,  halo model predictions are in good agreement with 
the bispectrum measured in numerical simulations.
However, the halo model predicts less than the observed 
configuration dependence of the 3PCF on $\sim$ Mpc scales.
This effect is mainly due to quadrupole moment deficit, possibly
related to the assumption of spherical halo geometry.
Our analysis shows that using its harmonic decomposition,
the full configuration dependence of the 3PCF in the non-linear regime 
can be compressed into just a few numbers, the lowest multipoles. 
Moreover, these multipoles are closely related to the highest
signal to noise eigenmodes of the 3PCF.
Therefore this estimator may simplify
future analyses aimed at constraining cosmological and
halo model parameters from observational data.

\end{abstract}

\keywords{cosmology: theory --- large-scale structure of universe --- methods:
statistical --- numerical}
\section{Introduction}

Galaxy formation and evolution still lacks a compelling explanation from 
first principles. 
In the absence of a successful theory, galaxy clustering can be described
assuming that galaxies are biased tracers of the underlying dark matter 
distribution. 
Recently, the halo model for gravitational clustering  
\citep{NeymanScott1952,Peebles1974,McClelland1977a,McClelland1977b, McClelland1978} 
has been revived in cosmology as an 
attempt to provide an accurate picture of gravitational clustering in the 
non-linear regime as seen by high-resolution N-body simulations 
\citep{ScherrerBertschinger1991,ShethJain1997,Seljak2000, MaFry2000a,MaFry2000b,ScoccimarroEtal2001, CoorayHu2001}.

Halo model provides a simple prescription for the analytic
computation of N-point correlation functions which are the most widely used
statistics for gravitational clustering (see \cite{CooraySheth2002} and references therein).
 In particular, \cite{Seljak2000} suggested that the power spectrum of dark matter and galaxies 
are consistent with this approach, whereas
\cite{PeacockSmith2000} reached similar conclusions using 
mock catalogs.  
\cite{MaFry2000a,MaFry2000b} derived predictions for the power spectrum, 
bispectrum and its Fourier counterparts, the two and three-point correlation functions,
that were found to be in agreement with numerical simulations.
\cite{ScoccimarroEtal2001} (hereafter SSHJ), 
discussed the halo model predictions for the power spectrum, and bispectrum, 
as well as the higher-order cumulants,
$S_p$, and concluded that dark matter clustering in the halo model is in good agreement 
with measurements from N-body simulations on large scales ($\R > 1 \mpc$).
However, the agreement was at the 20 \% level only on smaller 
scales ($\R \simlt 1 \mpc$). 
They also showed a first comparison with the variance from APM galaxies.

Progress on numerical simulations has triggered a number of
theoretical developments in the halo model. New elements such as 
substructure \cite[see \eg][]{Sheth2003,ShethJain2003,DolneyEtal2004} and halo geometry \citep{JingSuto2002} 
have been recently incorporated into the model to provide a more
realistic description of dark matter clustering.
Other developments have focused on describing galaxy bias using the so-called halo occupation distribution, 
a prescription of filling halos with galaxies in a stochastic
fashion \citep{BerlindWeinberg2002,BerlindEtal2003,ZehaviEtal2004}. 
 
Recently, \cite{TakadaJain2003a} have thoroughly explored halo model 
predictions for the 
three-point correlation function (3PCF hereafter) 
of dark matter and galaxies. 
They concluded that halo model predictions for the
dark matter described adequately only
certain triangle configurations. They fail
to match simulations for triangles with side 
length $\R \approx ~1 \mpc$. 
\cite{WangEtal2004}  investigated the two- and three-point
functions and compared analytic
results to dark matter simulations and galaxy clustering observations 
from the 2dF. They claim
their predictions closely reproduce numerical results 
and observations.   

In this paper we focus on the dark matter three-point correlation functions in 
real and Fourier space. In particular, our principal aim is to carry out a detailed comparison
of halo model predictions with measurements in high resolution simulations
to test the validity of the model and constrain halo model parameters.

The present analysis differs from previous work in the literature in 
several ways: 
\begin{itemize}

\item{
We present new technology to perform the
necessary integrals using multipole expansion, in particular
the dimension of the integrals is significantly reduced in
our prescription.}

\item{For the first time, we test the validity of the halo model 
against N-body simulations simultaneously in real and Fourier
space, at the two and three-point level.}
 
\item{We constrain halo parameters in a self-consistent fashion:
all statistics are predicted from halo model parameters. 
In particular, we do not use the \cite{SmithEtal2003} 
as an ingredient of the model, which has become common
usage when constraining bias from observational data \citep[see \eg][]{ZehaviEtal2004,WangEtal2004}.
We show that using the halo model in a self-consistent way provides a
 good fit to N-body simulations, provided one leaves the halo boundary 
as a free parameter, and large-mass haloes, not present in simulations, 
are removed from the mass function accordingly.}

\item{
We extend our analysis for the recently proposed multipole 
expansion of three-point statistics, which expresses rotation
invariance naturally. This allows the compression
three-point statistics  into a few 
multipoles even in the non-linear regime. This is 
especially convenient for constraining non-Gaussianity 
from gravitational clustering.}

\end{itemize}

\section{Halo Model}
\label{sec:model}

According to the halo model picture, 
the non-linearly evolved dark matter distribution is described 
in terms of the clustering properties dark haloes, 
quasi-equilibrium objects formed by gravitational collapse, 
\citep[see][for a review and references therein]{CooraySheth2002}. 
In this context, correlation functions are decomposed 
into contributions arising 
from correlations among particles inhabiting dark-matter haloes.
For convenience throughout this paper 
we shall closely follow SSHJ to 
describe the statistics of gravitational clustering within the halo model.

\subsection{Halo profile}
  
Numerical simulations suggest that dark haloes have a universal profile 
\citep{NavarroEtal1997, MooreEtal1998}. Here we shall adopt the NFW profile,
\be
\rho(r) = \frac{f}{4 \pi} {\left(\frac{c}{\rv} \right)}^3 \frac{1}{c r/\rv (1 + c r/\rv)}
\ee
and its Fourier transform, $\rho(k) = \int_0^\rv 4\pi r^2 dr \rho(r) j_0(k r)$, where $j_0(x) = \sin x/x$ 
is the zeroth-order spherical Bessel function,  
takes the form (see SSHJ),
\be
\rho(k,y) = f \left[ \sin \eta ~\{\rm{Si}[\eta(1+c)]-\rm{Si}(\eta)\} + \cos \eta ~\{\rm{Ci}[\eta(1+c)]-\rm{Ci}(\eta)\} 
- \frac{\sin(\eta c)}{\eta(1+c)} \right]
\ee
with $f = 1/\left[\ln(1+c)-c/(1+c)\right]$, $\eta = k \rv/c$, $c$ is the concentration parameter,
and $\rm{Si}$ and $\rm{Ci}$ are the sine and cosine integral functions. 
The virial radius, $\rv$, is the characteristic scale which separates the inner $\R^{-1}$ from the
outer $\R^{-3}$ behavior of the profile.
Above we have assumed that the halo profile is truncated at $\rv$, an assumption that shall be
relaxed in order to match clustering measurements in N-body simulations (see \S\ref{sec:N-body}).  
The concentration parameter is not well constrained from
simulations and we adopt the standard halo parametrization with halo mass
\be
c(M) = c_0 \left(\frac{M}{M_*}\right)^{-\beta}
\label{eq:cpar}
\ee
$M_*$ sets the non-linearity scale $\sigma(M_*) = \delta_c$, being $\sigma$ the linear rms mass fluctuation
and $\delta_c = 1.686$ is the linear overdensity
required for spherical collapse. We shall assume the values suggested 
by numerical simulations $c_0 = 9$, $\beta = 0.13$ \cite{BullockEtal2001},
but have checked that changing halo concentration parameters by as much as 50 \% does 
not affect significantly our predictions within the range of scales probed by our simulations.

\subsection{Halo mass function}

A basic ingredient of the halo model is the mass function, $n(M)$, which describes 
how many objects of mass in the range $M$ and $M+dM$ end up collapsing to form bound structures.
Here we use the extension of the Press-Schechter formalism provided by \cite{ShethTormen1999}
that accurately describes N-body results \citep{JenkinsEtal2001}:
\be
n(M) M dM = \bar{\rho} \frac{dy}{y} n(\nu) = \bar{\rho} \frac{dy}{y} A~\gamma \sqrt{\frac{g(\nu)}{2\pi}} \left(1+g(\nu)^{-p}\right) \exp(-g(\nu)/2)
\ee
with $\gamma = d\ln \sigma^2/d\ln R$, $g(\nu) = \alpha \nu^2$, $\alpha = 0.707$, $\nu = \delta_c/\sigma$,  
$A=0.322$, $p=0.3$, $y = (R/R_*) = (M/M_*)^{1/3}$, $R$ is the lagrangian radius, and  
$R_* = \rv ~\Delta$ is the non-linear (lagrangian) scale,
where $\Delta = 200, 340$ for $\Omega_m = 1, 0.3$ cosmologies, respectively.

\subsection{Halo clustering}

Halo correlations can be included in the model 
by assuming a biasing prescription between the halo and the underlying mass distribution.
We follow the approach introduced by \cite{MoWhite1996} in the context of the spherical collapse model,
and adopt the fitting formula given by \cite{ShethTormen1999}:
\be
b(\nu) = 1 + \frac{g(\nu)-1}{\delta_c} + \frac{2 p}{\delta_c(1+g(\nu)^p)}
\ee
and we neglect quadratic and higher-order biasing terms.
We note that a non-vanishing quadratic bias only affects the three-point statistics (or higher orders)
on large scales and we have checked that its contribution is a few per cent at most.

\section{Halo Model Statistics}

Halo model statistics are particularly simple to formulate in Fourier space.
Complications arising from convolutions of halo profiles become
simple products in transform space. 

According to the halo model approach, the non-linear power spectrum of the mass
fluctuations is the result of two independent contributions: 
one coming from the single halo profile auto-correlation that dominates on small-scales and 
another one given by the correlation among dark matter particles in different halos  
that accounts for the large-scale clustering.
This way we can write,
\be
P(k) = \left[I_{11}(k)\right]^2~P_L(k)+ I_{02}(k,k)
\ee
where,
\be
I_{ij}(k_1,\cdots,k_j) = \int \frac{dy}{y} n(y) b_i(y)\left[\rho(k_1,y)\cdots \rho(k_j,y)\right]
{\left(\frac{R^3_* y^3}{6\pi^2}\right)}^{j-1}
\ee
where $b_0 = 1$, $b1 = b(\nu)$, and $b_i = 0$ for $i>1$ as we neglect quadratic and higher-order
biasing terms, and $P_L(k)$ is the linear power spectrum, for which we assume the 
\cite{BondEfstathiou1984} parametrization. 
The two-point correlation function follows by Fourier transforming $P(k)$,
\be
\xi(r) = \int \frac{k^2}{2\pi^2} dk~P(k) ~j_0(k r)
\ee
where $j_0$ is the zero-order spherical Bessel function.
 
The bispectrum, $B_{123} \equiv B(k_1,k_2,k_3)$, can be expressed as 
a sum of three-point correlations among mass particles 
residing in one, two, and three haloes, respectively:
\be
B_{123} = \left[\prod_{i=1}^3 I_{11}(k_i)\right]B^{PT}_{123} + 
                 \{I_{11}(k_1)I_{12}(k_2,k_3)P_L(k_1) + { \rm perm(1,2,3)}\} +
                 I_{03}(k_1,k_2,k_3)
\ee
where $B^{PT}$ denotes the bispectrum 
from second order Perturbation Theory (PT).
By design, halo model recovers weakly non-linear 
theory predictions
on large scales, since $k\rightarrow 0$, $I_{11}\approx 1$
and $I_{12},I_{02},I_{03} \approx 0$, 
or equivalently, $P\approx P_L$, $B_{123} \approx B^{PT}$.

The 3PCF is a triple  Fourier transform of the bispectrum,
\be
\zeta (r_1,r_2,r_3) = \int \prod_{i=1}^3 d^3{\bf k}_i~B_{123}
~e^{i({\bf k}_1\cdot{\bf r}_1+{\bf k}_2\cdot {\bf r}_2+{\bf k}_3\cdot {\bf r}_3)}\delta_{Dirac}({\bf k}_1+{\bf k}_2+{\bf k}_3).
\ee
In the context of halo models, $\zeta$ is a complex
object: \cite{TakadaJain2003a} have shown that the evaluation of the 
two- and three-halo terms in real space involves 
eight- and twelve-dimensional integrals, respectively; 
this is untractable. They reduce the dimensionality
of the integrals to two by switching to Fourier-space and 
taking a number of approximations in the corresponding kernels.

Our approach is significantly different in that, although we
also use Fourier-space formalism, we further
decompose correlation functions into harmonic multipoles.
This greatly simplifies the exact analytic expressions,
therefore we do not need to take the approximations used by \cite{TakadaJain2003a}
for the 2- and 3-halo terms. We note that, for the 1-halo term, their 
estimator is expressed as a four dimensional integration and it is exact.
The only approximation we take is that we use a finite number of
multipoles. We will see that this is an excellent approximation, except for 
degenerate (isosceles) triangles, where the amplitude of the harmonic coefficients 
decrease slowly with multipole order.
In our formalism, the 3PCF can be expressed
as a two-dimensional Hankel transform of the bispectrum, 
for each harmonic multipole \citep{Szapudi2004}:
\be
\zeta (r_1,r_2,\theta) = \sum_{\ell} \frac{2\ell+1}{4\pi} \zeta_{\ell}(r_1,r_2)~P_{\ell}(\cos\theta)
\label{eq:zeta}
\ee
where $\theta$ is the angle between ${\bf {r_1}}$ and ${\bf {r_2}}$, $P_{\ell}$ is the legendre polynomial of order $\ell$, $\zeta_{\ell} (r_1,r_2)$ is 
the harmonic 3PCF, 
\be
\zeta_{\ell} (r_1,r_2) = \int dk_1~dk_2\frac{k_1^2}{2 \pi^2}\frac{k_2^2}{2\pi^2}(-1)^{\ell}~B_{\ell}(k_1,k_2)j_{\ell}(k_1 r_1) j_{\ell}(k_2 r_2)
\label{eq:zetal}
\ee
and $B_{\ell}$ is the Legendre transform of the bispectrum,
\be
B_{\ell}(k_1,k_2) = 2\pi \int_{-1}^{1} d(\cos\theta)~B(k_1,k_2,\theta)~P_{\ell}(\cos\theta) \ .
\label{eq:bl}
\ee
High multipoles ($\ell\gg 1$), only important for the
isosceles triangles,  can be easily computed employing the 
large-$\ell$ limit of spherical
Bessel functions, 
$j_{\ell}(x) \approx \sqrt{\pi/(2\ell+1)}~\delta_{Dirac}(\ell+1/2-x)$,
\be
\zeta_{\ell} (r_1,r_2) = \left(\frac{L}{2\pi}\right)^3\frac{1}{(r_1r_2)^3}(-1)^{\ell} B_{\ell}\left(\frac{L}{r_1},\frac{L}{r_2}\right)
\ee
where $L \equiv \ell+1/2$. 
It is convenient to define the reduced (or hierarchical) 3PCF,
\be
Q(r_1,r_2,r_3) =  \frac{\zeta(r_1,r_2,r_3)}{\xi(r_1)\xi(r_2)+\xi(r_2)\xi(r_3)+\xi(r_3)\xi(r_1)}
\label{eq:q}
\ee
and similarly, the reduced bispectrum in Fourier space,
\be
Q(k_1,k_2,k_3) = \frac{B_{123}}{{P(k_1)P(k_2) + P(k_2)P(k_3)+ P(k_3)P(k_1)}}.
\label{eq:qk}
\ee
PT and the stable clustering hypothesis predict them to be 
weakly dependent of scale 
in the quasi-linear and non-linear regimes, respectively.
Similar to eq(\ref{eq:bl}), it is convenient to decompose the 
reduced 3PCF, $Q(r_1,r_2,r_3)$, in its harmonic 
multipoles by taking its Legendre transform,
\be
Q_{\ell} (r_1,r_2)= 2\pi \int_{-1}^{1} d(\cos\theta)~Q(r_1,r_2,\theta)~P_{\ell}(\cos\theta) 
\label{eq:ql}
\ee
where $\theta$ is the angle between ${\bf r_1}$ and ${\bf r_2}$. Analogous expressions
can be written for the harmonic multipoles of the reduced bispectrum.

Since the bispectrum itself is obtained by performing a one-dimensional 
integration of the halo correlations over the mass function, 
in order to compute the three-point function 
multipoles $\zeta_{\ell}$, as given by eq($\ref{eq:zetal}$),
four-dimensional integrals are required for each $\ell$
\footnote{the four integrals correspond to a mass function integral, Legendre transform of the bispectrum, and a double Bessel integral to get from Fourier to real space}.
Figure \ref{fig:q_lterms} shows how the 3PCF in the non-linear regime 
can be efficiently reconstructed from its lowest harmonic multipoles.
In particular, for the isosceles triangle with side length $r=1 ~\mpc$, 
only three harmonics contribute significantly to the 3PCF. 
The monopole essentially determines the amplitude, 
the (negative) quadrupole and, to a lesser extent, 
the octopole shape the configuration dependence 
(see top left and middle panels).
Thus the full configuration dependence of the 3PCF can be
reconstructed from these three harmonics (see top right panel) to
an excellent approximation. 
Only collinear configurations, $\theta = 0, \pi$ have 
(increasingly smaller) contributions from higher order terms. 
Similarly, the triangles with $r_1 = 2~ \mpc$, and 
$r_2 = 6~ \mpc$ are shaped 
by the quadrupole and therefore, the configuration dependence is 
encoded in this single harmonic to a large extent (see lower panels).
These examples are typical for all non-degenerate triangles we have checked.

As we shall show below (see \S\ref{sec:N-body}), in general,  
{\em only the lowest multipoles are non-zero 
and thus full configuration dependence of the
 non-linear cosmological 3PCF can be compressed in a few numbers.} 
Thus using the multipole expansion greatly reduces
the problem of constraining non-gaussianity from gravitational 
clustering.

\section{Comparison to Numerical Simulations}
\label{sec:N-body}

We have used two sets of \lcdm simulations from the public Virgo simulation 
archive\footnote{http://www.mpa-garching.mpg.de/Virgo} 
with cosmological parameters 
$\Omega_0=0.3$, $\Omega_\Lambda=0.7$, $h=0.7$, $\Gamma=0.21$ 
and $\sigma_8=0.9$ and no baryons.  The original Virgo simulation has $256^3$ 
particles in a box-size of 
$L=239.5h^{-1}$Mpc, mass resolution of $6.86\cdot10^{10} h^{-1}M_{\sun}$ and softening length
$L_{soft} = 25 h^{-1}$kpc, 
and a larger box (VLS) simulation $L=479h^{-1}$Mpc
that contains $512^3$ particles, same mass resolution than the original Virgo simulation and 
$L_{soft} = 30 h^{-1}$kpc .
These simulations have been gravitationally evolved using a P3M code
\citep{FarlandEtal1998, CouchmanEtal1995}.  

Bispectrum is computed through FFT's on $256^3$ grid-points  with the method of 
\cite{ScoccimarroEtal1998} (see their Appendix A).
The 3PCF is measured with the fast algorithm based on
multi-resolution KD-trees of \cite{MooreEtal2001} and \cite{GrayEtal2004}
using the estimator of \cite{SzapudiSzalay1998}. 
Limited by computational resources, we dilute simulations to 10 percent for
the 3PCF at scales $\R \simlt 2 ~\mpc$ and at 1 percent at larger scales.
For these computations the number of points in auxiliary random sets are roughly 10 times 
larger than those of diluted simulation data sets.  
The large box simulation
is cut into eight independent sub-volumes (octants) of half box size and 
measured separately.
Mean values are obtained by averaging these eight sub-volumes and 
the orginal Virgo simulation.
Errorbars are simply computed from the dispersion over sub-volumes.
We note that this error estimate might underestimate {\it true} 
errors\footnote{Using a large number of very large box simulacions 
should yield a more correct error estimate on large scales} since
sub-volumes are somewhat correlated. However, for the small scales analyzed in this
paper, our sub-samples are effectively independent.

Our approach is to discuss halo model predictions for a  
{\it fiducial} model that provides a good
fit to the 2-point statistics (\ie power spectrum and 2PCF) 
measured in the simulations. Then we explore systematically
the ability of such a model of describing 3-point statistics in N-body experiments.
We set cosmological parameters to match those of our simulations and 
the halo concentration parameter is set to $c_0 = 9$ and 
$\beta = 0.13$ in eq($\ref{eq:cpar}$), as in
SSHJ. We have verified that changing the amplitude and slope
of the concentration parameter within a reasonable range 
(\ie by a factor $\sim 2$)
has a negligible effect on the 2PCF and 3PCF for  
$k \simlt 10$, or ${\R} \simgt 0.2 \mpc$  
\citep[see also Fig.2 in][]{TakadaJain2003a}, therefore we fix
this parameter to its fiducial values to simplify further studies.
 
Basic halo model parameters yield a 2PCF that does not fit well our N-body results.
This is clearly seen in Figs.$\ref{fig:xi2_paper_hb}$-$\ref{fig:qeqk_paper_mc}$,
especially for scales, $\R \approx 1~\mpc$.   
To improve the precision, we introduce standard ``tweaks''
of the basic set of ingredients of the model.
The assumption that the halo boundary 
(the radius up to which halo mass is encompassed) 
is exactly given by the virial radius is in fact quite arbitrary.
Treating this as an additional free parameter to the 1-halo terms 
in the 2PCF and 3PCF improves the 
{\it transition scales}, $0.5-5~\mpc$, 
between linear (or quasi-linear) to fully non-linear scales
(see Figs.$\ref{fig:xi2_paper_hb}$ and $\ref{fig:qeqk_paper_hb}$).
Fig.$\ref{fig:qeqk_paper_hb}$ shows that a
halo boundary beyond the virial radius increases the 
2PCF on small scales and 
reduces the ``bump'' in the reduced bispectrum 
for equilateral triangles $Q_{eq}(k)$. 
Setting the boundary to the fiducial value, $1.3\rv$, 
yields good agreement both with the 2PCF and 3PCF in simulations.
In particular, for the 2PCF, the agreement with N-body is comparable to the 
fitting function  provided by Smith et al.(2003). 
However, on large scales, halo model tends to slightly overpredict
simulations, whereas Smith et al. slightly underpredicts them.
Similarly, Fig.$\ref{fig:pk}$ shows that our fiducial model provides a good fit to 
the matter power spectrum measured in our N-body simulations.

Our results are in full agreement with that of \cite{TakadaJain2003a} 
who also implemented the  ``exclusion
effect''. This excludes correlations between haloes at 
distances smaller than the
sum of their virial radii. Halo exclusion is an inherently
real space phenomenon, and it would be difficult, if not impossible,
to correctly implement in our Fourier-based approach. We have
checked using approximations (which essentially cut off the
two-halo term on small scales) 
that halo exclusion plays a subdominant role in describing the
3PCF \citep[see also Fig.8 in][]{TakadaJain2003a}.

In addition, we try to mimic the finite volume effects 
affecting N-body simulations by imposing
a cut-off in the mass function, 
excluding haloes of masses $M \simgt M_{cut}$. 
This has a critical effect on clustering on
scales $\R \simlt 5 \mpc$. We find that imposing 
no cut-off overpredicts the  clustering observed in simulations
whereas excluding haloes of $\M \simgt 10^{14.5}\Msun$ 
results in too low a 2PCF. This is even
more evident in the reduced bispectrum for equilateral triangles $Q_{eq}(k)$ 
(see Fig.$\ref{fig:qeqk_paper_hb}$). 
Our fiducial value $\M_{cut} = 10^{15}\Msun$ reproduces
reasonably well simulation results. 
This behavior was already observed by SSHJ and
\cite{WangEtal2004} who found different best-fit 
values for their smaller-box ($L = 300 \mpc$)
simulations. While these works did not discuss the effect
of the  mass cut-off on the 2PCF, we have made sure
that our fiducial choice, motivated by the 3PCF,
still reproduces the 2PCF accurately.

The above two extra parameters fix our fiducial model.  
In what follows, we shall use it
to work out detailed predictions for the 3PCF and its harmonic multipoles
$\zeta_{\ell}$ (see eqs($\ref{eq:zeta}$) and ($\ref{eq:zetal}$)), and compare
them to measurements N-body simulations.
Fig.$\ref{fig:q3k_scales}$ displays the reduced bispectrum for 
several different triangular configurations.
It appears that large scales present a more 
pronounced configuration dependence than predicted by the three-halo term.
Triangles on small-scales are rather flat, except for 
the ``shoulder'' displayed by the isosceles triangles. 

Halo model prediction for the reduced 3PCF for 
equilateral triangles, $Q_{eq}(r)$
is in good agreement with N-body results 
as shown in  Fig.$\ref{fig:qeq_paper}$. 
On scales $\R ~\simgt 3 \mpc$ the model slightly overpredicts
simulations, similarly to the 2PCF on the same scales 
(see Fig.$\ref{fig:xi2_paper_hb}$ and Fig.$\ref{fig:xi2_paper_mc}$).
Note that the non-linear halo model
converges, albeit slowly, to the PT on large-scales.

Fig.$\ref{fig:zeta_scales}$ displays the configuration 
dependence of the 3PCF for the {\it transition scales} as
predicted by the halo model (lines) compared 
to numerical simulations (symbols).
The agreement between the model and simulations
is generally within the errorbars. Moreover, 
the rich configuration and scale dependence
made up by the non-trivial 
interplay between different halo terms 
is  observed in simulations. This can be considered as
the best confirmation that the basic idea of halo models
carries over to the three-point function without major modifications.

Fig.$\ref{fig:q3_scales}$  shows the reduced 3PCF $Q(r)$ for the
same triangles whereas Fig.$\ref{fig:q3_3d_sims_r2p0}$
shows an interpolated surface displaying the $(\theta,q)$ dependence 
for the mean values of the N-body for $\R = 2 ~\mpc$.
Note that since $Q(r)$ is predicted to be of order unity 
for all configurations, it is more suitable for high-precision 
comparisons on linear scale plots. On the down side,  the
complex structure of this ratio statistic 
combining 2PCF and 3PCF complicates
interpretation. In particular, halo model describes the 
configuration dependence of the reduced 
3PCF accurately for $\R \simlt 0.5 \mpc$, 
but it becomes increasingly inaccurate for larger scales.
The slight overall amplitude mismatch observed 
(see \eg the case for isosceles triangles, $q=1$, 
in left column of Fig.$\ref{fig:q3_scales}$) is due to the halo model 2PCF 
overpredicting simulations. 
We have checked that using fitting formula of
\citep{SmithEtal2003} for the 2PCF instead of the halo model
does not  improve the agreement with simulations significantly.
Note that for intermediate angles theoretical predictions always agree 
with the simulations, in line with our earlier results 
for equilateral triangles  (see Fig.$\ref{fig:qeq_paper}$).
The most visible difference between model and N-body results 
is the lack of configuration
dependence predicted by the halo model on larger scales, $\R ~\simgt 0.5 \mpc$.

The multipole expansion of the 3PCF, eq.(\ref{eq:zetal}), 
is particularly useful to 
pinpoint why the halo model fails to reproduce numerical results.
Figs.$\ref{fig:zl_0p25_comb}$-$\ref{fig:zl_2p0_comb}$ 
show the 3PCF multipoles for different
triangles and scales $\R = 0.25-2 ~\mpc$. In general, 
the amplitude of the coefficients 
falls off rapidly with multipole order. This makes multipoles
especially convenient to compress  information in the
cosmological 3PCF in the non-linear regime \citep[see also][]{Szapudi2004}.

Isosceles triangles are exceptional: on small angles the third side
of the triangle is very small, which results in a rapid increase.
This can be described in multipole space only with a high number
of multipoles (such as a Dirac $\delta$-function would have infinite
number of multipoles). The situation worsens towards larger scales,
due to even more rapid increase of the 3PCF on small angles.
While this technical deficiency can be overcome by simply
smoothing the correlation function over angles (``band limiting''),
we note that on scales above $\R ~\simgt 5 \mpc$, where
this effect becomes severe, PT becomes more and more accurate, 
therefore halo models are not necessary for dark matter predictions
\citep{Fry1984, JingBorner1997, BarrigaGaztanaga2002}.

More interestingly, for $\R ~\simgt 0.5 \mpc$, halo model 
consistently underpredicts 
the quadrupole moment (and to a lesser extent, the octopole).
{\it The observed 
lack of configuration dependence in the halo model with respect to simulations 
is primarily due to a quadrupole deficit in the prediction} 
(see Figs.$\ref{fig:zeta_scales}$ and $\ref{fig:q3_scales}$). 
This might be caused by the 2-halo term, 
which shows hardly any variation with angle on 
large-scales $\R ~\simgt 0.5 \mpc$
(see Figs.$\ref{fig:zeta_scales}$ and $\ref{fig:q3_scales}$), 
contrary to what one would expect for halo-halo spatial correlations.  
One possible reason for this quadrupole deficit in the two-halo term  
is that our implementation of the halo model assumes 
that haloes are spherical, whereas
real virialized objects in high-resolution 
CDM simulations do have a typically asymmetric (triaxial)
shape \citep[see \eg][]{BarnesEfstathiou1987, JingSuto2002, MooreEtal2004}.
SSHJ already pointed out in their bispectrum analysis
that relaxing the sphericity hypothesis 
for halo shapes could bring halo model 
to a closer agreement with the configuration
dependence observed in numerical simulations. 
Although our analysis suggests a similar conclusion, 
it is unclear whether one-halo
(as pointed out by SSHJ) or rather two-halo terms (that show 
a significant lack of configuration dependence at all scales) 
should carry the missing quadrupole.

On the other hand halo substructure \citep{Sheth2003, ShethJain2003} 
could play a significant role on scales comparable 
to large-cluster sized haloes ~$1\mpc$.
However, a recent analysis \citep{DolneyEtal2004} suggests 
that substructure tends to attenuate the amplitude 
of the reduced bispectrum on small scales. This 
would render our analytic predictions, at least our fiducial model,
in disagreement with N-body measurements 
(see Figs. \ref{fig:qeqk_paper_hb} and \ref{fig:qeqk_paper_mc}).

Clearly, there are other potential improvements to 
our implementation of the halo model, most importantly a 
consistent definition of the mass function for 
a modified halo boundary \citep[see][]{White2002,HuKravtsov2003},
or allowing for a steeper inner halo profile \citep[see][]{MooreEtal1998} 
what could bring model predictions closer to measurements in high-resolution 
N-body simulations. We plan to address these issues in future work.

\section{Conclusions}

The revival of the halo model in recent years has been triggered by the
ability of high-resolution N-body simulations to test theoretical predictions
with great precision. 
In this paper we have presented a detailed comparison of 
halo model predictions against
simulations for the 2PCF and 3PCF in real and Fourier space.
Our analysis has focused on the transition scales, $\R \approx 0.25-5 \mpc$
that connect large (quasi-linear) scales, appropriately described by PT, with 
highly non-linear scales where phenomena 
such as the stable clustering hypothesis 
require higher resolution simulations to be accurately tested. 
 
Our results show that halo boundary and mass function 
cut-off have a significant effect
on the three-point correlation functions on these transition scales, 
and thus these statistics can be used to constrain such halo model parameters.
A fiducial model with halo boundary $\R = 1.3 \rv~\mpc$ and 
mass cut-off $M_{cut} = 10^{15}~\Msun$  
brings theoretical predictions in close agreement to our simulations.
The success of our fiducial model in explaining 
non-linear gravitational clustering
has been comprehensively demonstrated in Fourier space 
for the power spectrum (Fig.$\ref{fig:pk}$),
the reduced bispectrum (Figs.$\ref{fig:qeqk_paper_hb}$ and
$\ref{fig:qeqk_paper_mc}$), as well as in real space, through the 
two-point correlation function (Figs.$\ref{fig:xi2_paper_hb}$ and 
$\ref{fig:xi2_paper_mc}$), 3PCF (Fig.$\ref{fig:zeta_scales}$), 
and the reduced 3PCF
(Figs.$\ref{fig:qeq_paper}$, and $\ref{fig:q3_scales}$) along with its 
multipole moments
(Figs.$\ref{fig:zl_0p25_comb}$-$\ref{fig:zl_2p0_comb}$).

Halo model predictions are in good agreement with 
the configuration and scale dependence of the reduced bispectrum at
all the scales measured in N-body results, $k = 0.25-1~\impc$. 
On the other hand, while the model correctly predicts the amplitudes 
of the 3PCF on intermediate 
angles for all the scales tested, 
it exhibits a lack of configuration dependence that
becomes more significant on scales larger 
than the largest halos seen in simulations $\R \simgt 1 ~\mpc$. 
Although the reason for this is not clear, 
we suggest that non-spherical haloes 
could produce a more pronounced configuration dependence in theoretical
predictions. This could be realized through the two-halo terms 
that show a rather flat behavior in
the current implementation of the model. 
However if the orientation of the two halos is random, the non-sphericity
should cancel out.  
This issue certainly deserves further attention in future 
analyses of the dark-matter and galaxy N-point correlation functions. 

\begin{figure}[htb]
\figurenum{1}
\epsscale{1.}
\plotone{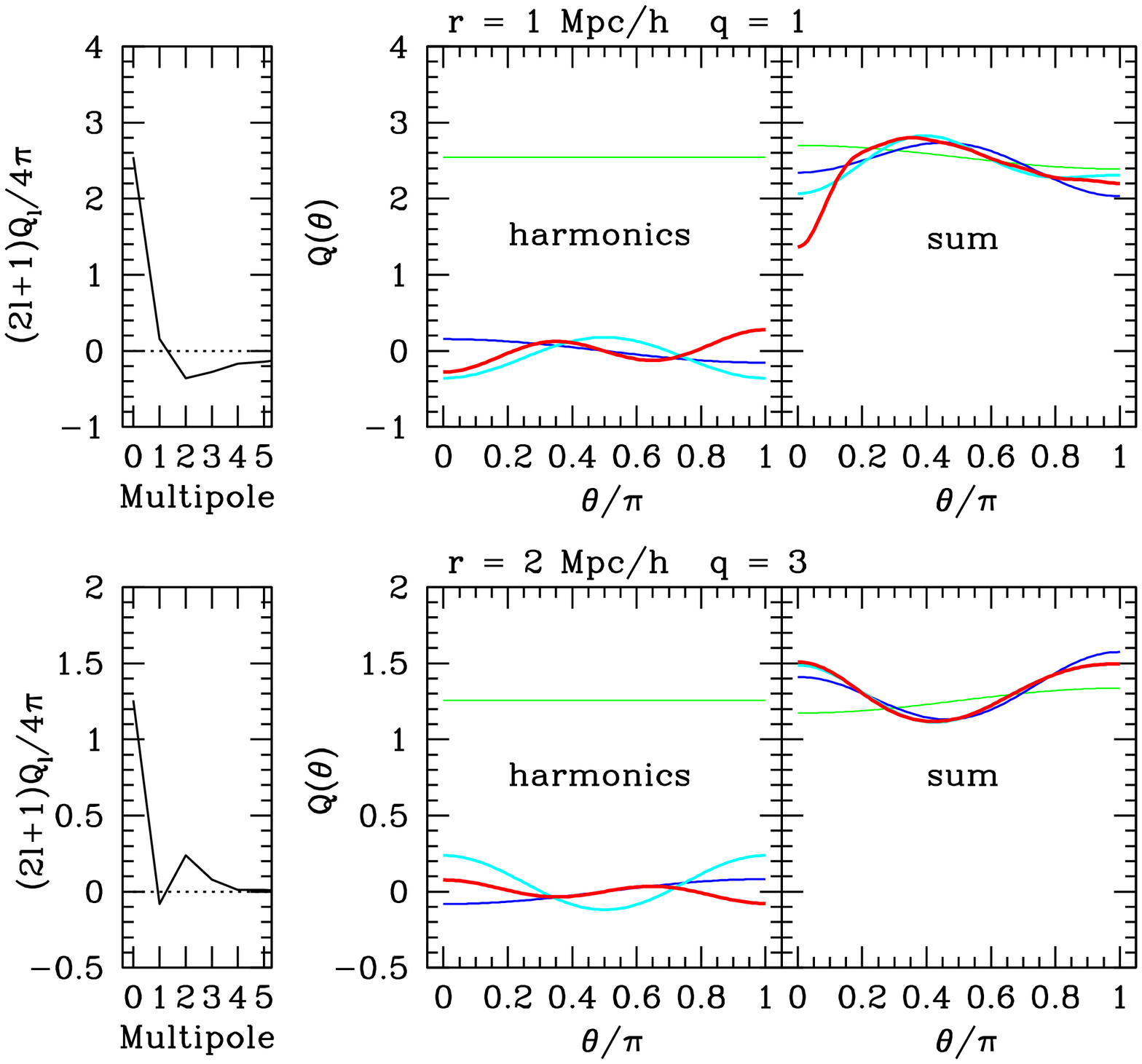}
\caption{Reconstructing the 3PCF form its harmonics.
Only the lowest harmonic coefficients are non-zero (see left panels).
 Middle panels show the configuration dependence of individual harmonic 
terms \ie harmonic coefficients times the Legendre polynomials. 
We display terms $\ell = 0,1,2$ and $3$ using increasingly thicker lines. 
Right panels show the sum over harmonics up to
a given $\ell_{max}  = 1,2,3$ and $10$ (from thin to thick lines).  
In general, the non-linear 3PCF
is dominated by the monopole, that sets the amplitude, and quadrupole, 
that shapes the configuration dependence. Top panels show isosceles 
triangles of length $r = 1~ \mpc$, while lower panels display more 
elongated triangles with basic length $r = 2~ \mpc$ and length-ratio $q = 3$. 
\label{fig:q_lterms}}
\end{figure}

\begin{figure}[htb]
\figurenum{2}
\epsscale{1.}
\plotone{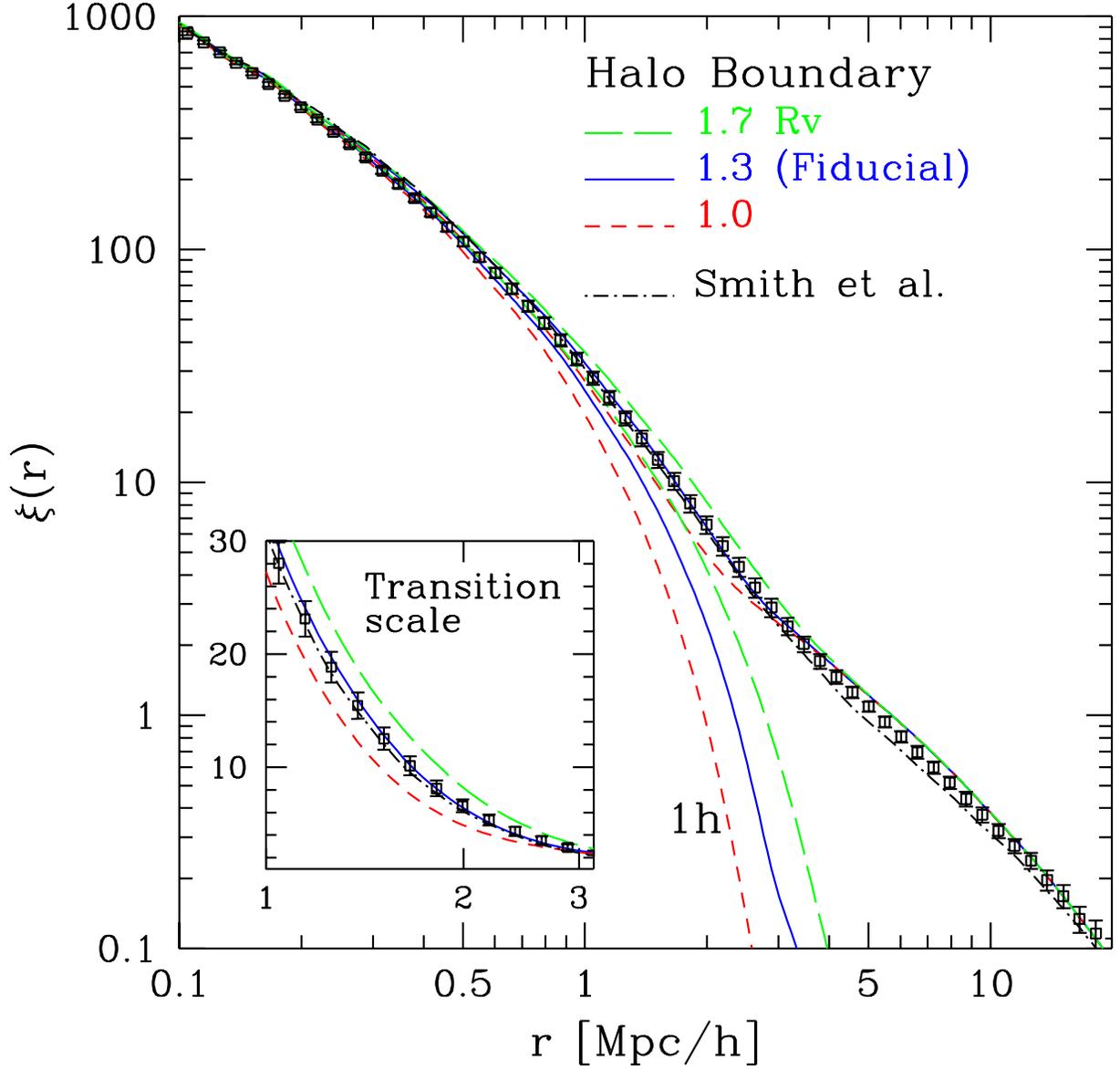}
\caption{Halo boundary effects on the 2PCF: 
defining halo boundary at $\R = \rv$ underestimates
$\xi(r)$ on the transition scales. Larger boundaries yield a 
larger one-halo (1h) contribution.
For $\R = 1.3 ~\rv$ (solid line), we find good agreement with N-body results. 
The inset shows a blow-up of the
transitions scales, where our fiducial model agrees well 
with numerical results and the 
\cite{SmithEtal2003} fitting function. On larger scales the fiducial 
model overpredicts numerical simulations, but
converges to them for $\R \simgt 20 \mpc$.
\label{fig:xi2_paper_hb}}
\end{figure}

\begin{figure}[htb]
\figurenum{3}
\epsscale{1.}
\plotone{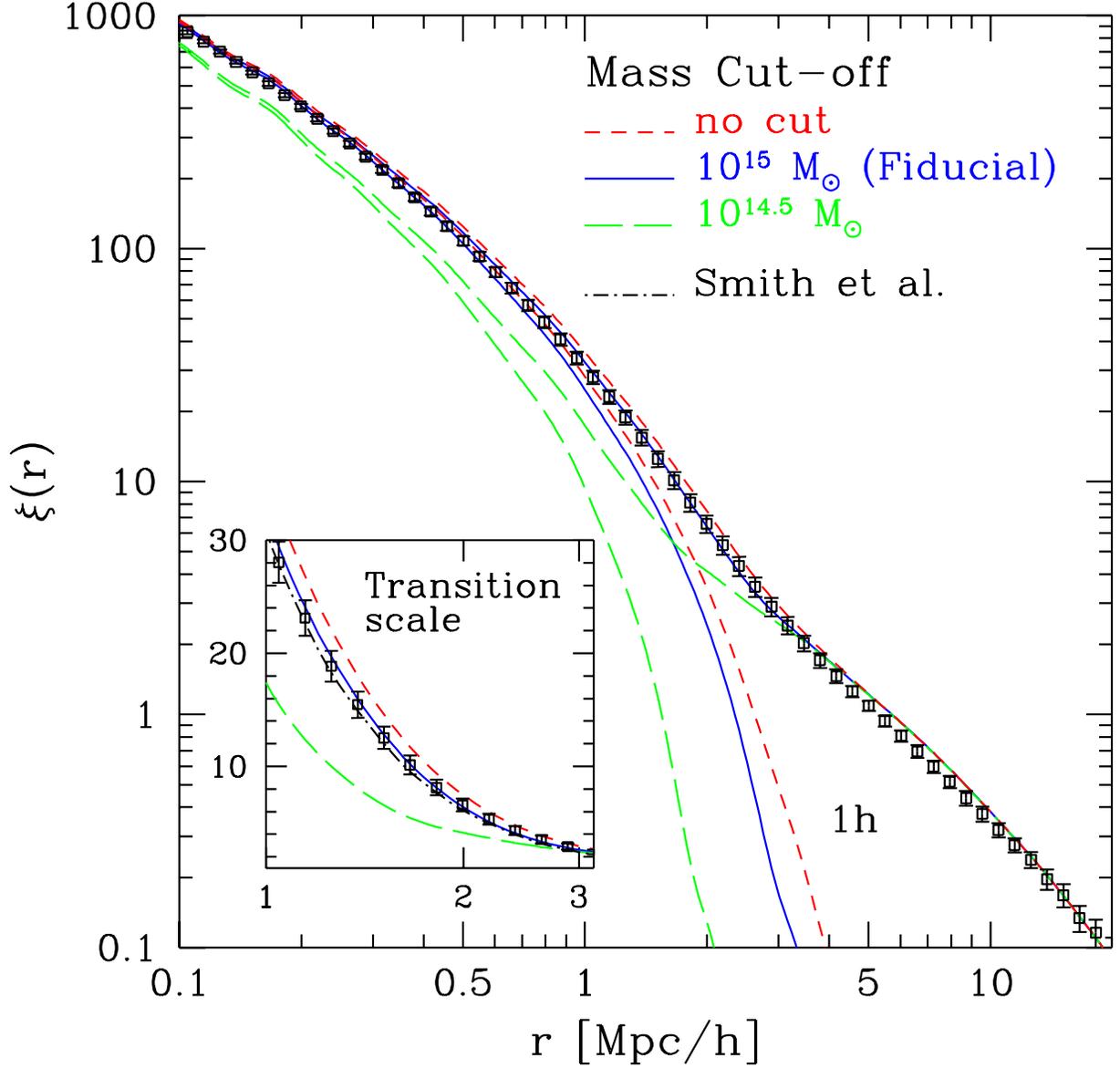}
\caption{Same as Fig.$\ref{fig:xi2_paper_hb}$ for the mass cut-off: 
in order to mimic the absence of haloes
of mass $M > M_{cut}$ in simulations, our fiducial model 
incorporates a cut-off in the 
mass function, $M_{cut} = 10^{15}\Msun$ (solid line), 
what lowers the 2PCF on non-linear scales 
(\ie those dominated by the one-halo term). 
The Smith et al. fitting function is only shown in the inset
for clarity.
\label{fig:xi2_paper_mc}}
\end{figure}

\begin{figure}[htb]
\figurenum{4}
\epsscale{1.}
\plotone{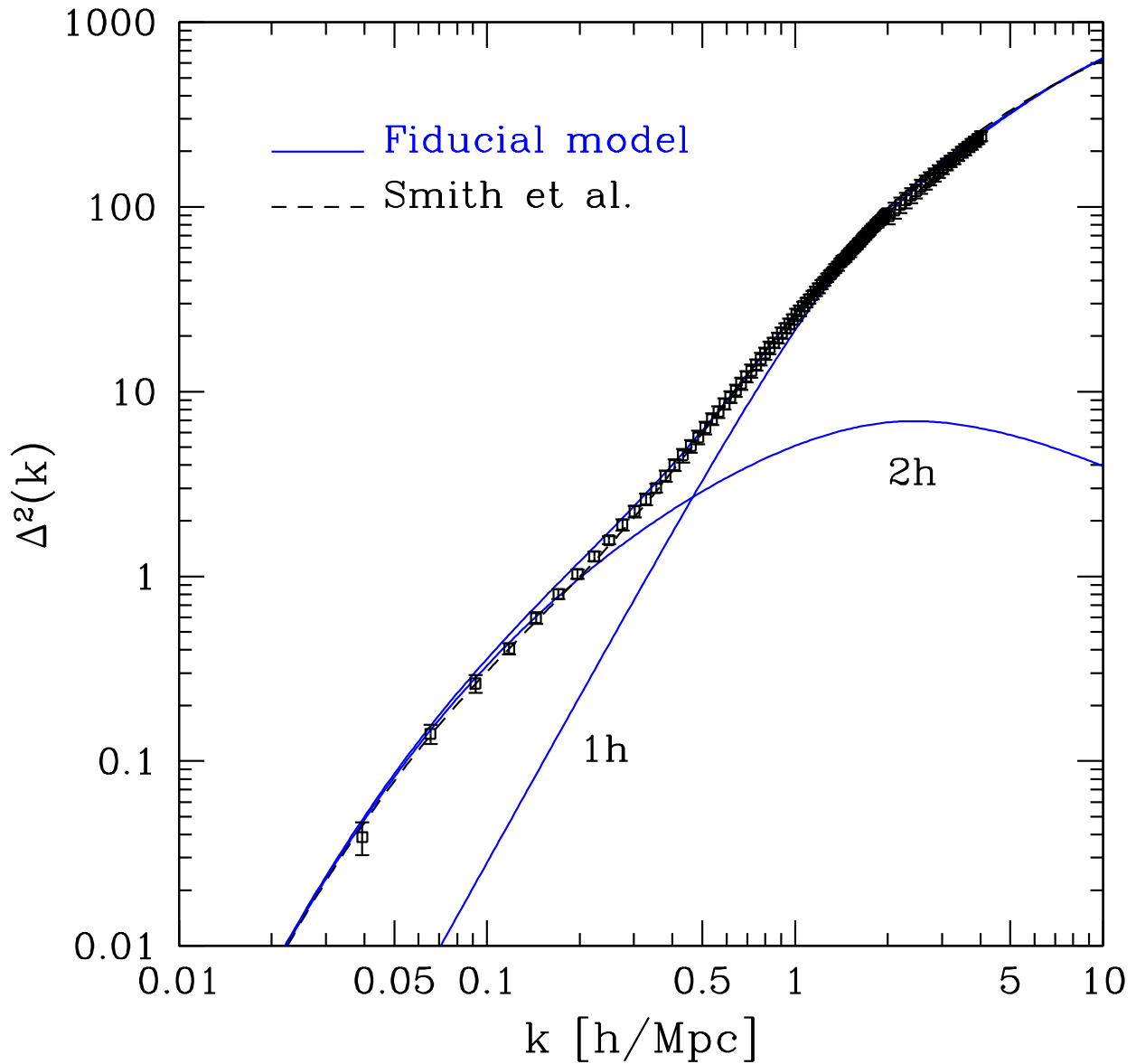}
\caption{Mass power-spectrum per logarithmic wavenumber, $\Delta^2(k) = k^3P(k)/2\pi^2$, 
in the fiducial halo model (solid line, see text for details) 
compared to N-body simulations (symbols). 
One (1h) and two (2h) halo contributions are shown 
to have comparable amplitude in the {\it transition scales} 
$k \approx 0.5 ~\impc$. For reference, it is also shown a
fitting function to numerical simulations (dashed line; \cite{SmithEtal2003}). 
Halo model agrees very well with N-body except on scales 
$k \approx 0.1-0.5 ~\impc$,  where it tends to slightly 
overpredict numerical simulations.  Note that we use a box-size  
$L=239.5~\mpc$ for small scales,  $k \simgt 2~\impc$, whereas a larger box (VLS) simulation $L=479~\mpc$
is used for larger scales, $k  \simlt 2 ~\impc$.  
\label{fig:pk}}
\end{figure}

\begin{figure}[htb]
\figurenum{5}
\epsscale{1.}
\plotone{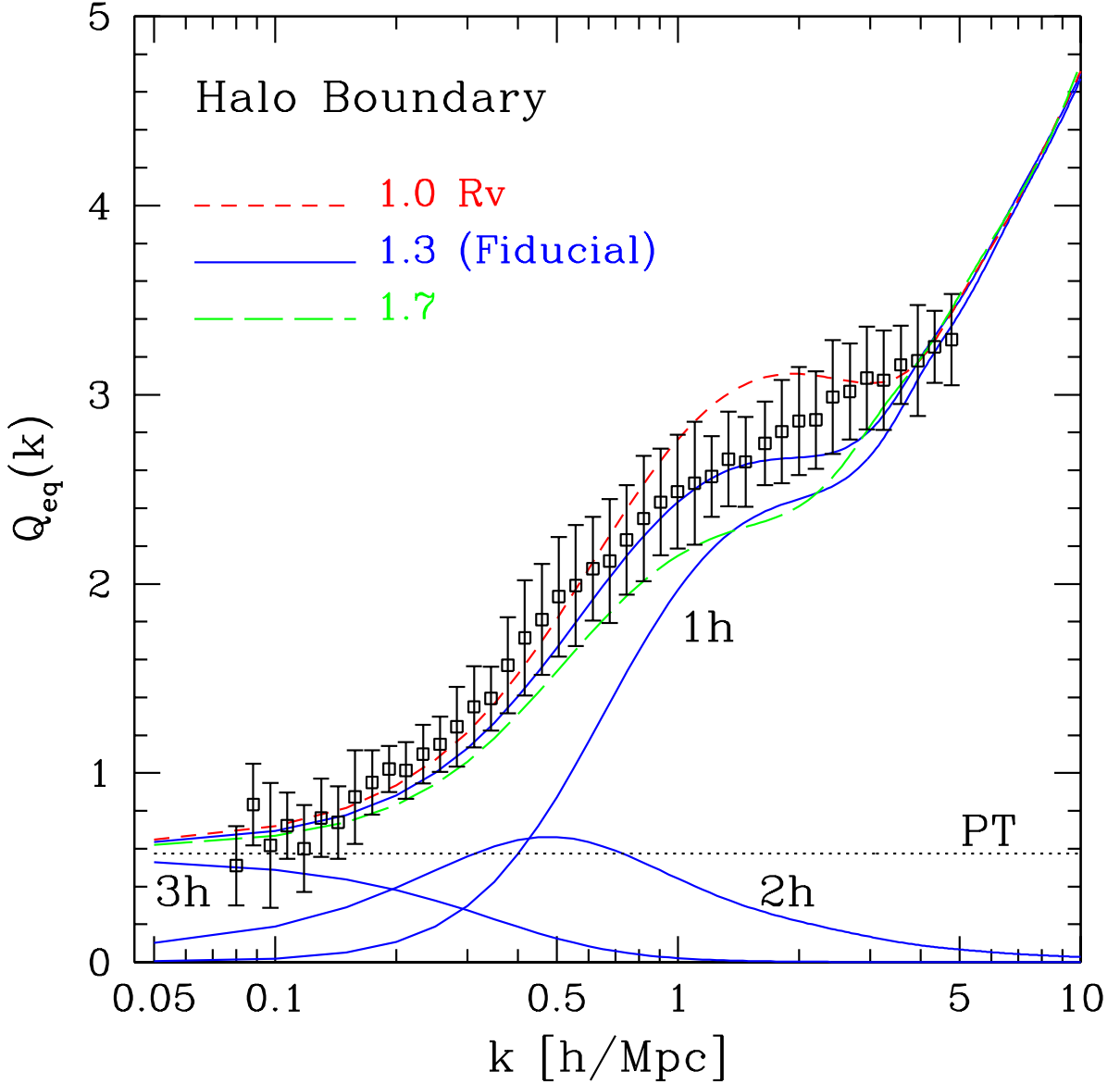}
\caption{Halo boundary effect on the reduced bispectrum for 
equilateral triangles $Q_{eq}(r)$.
$Q_{eq}(r)$ is very sensitive to changes in the halo boundary: 
larger boundaries enhance 
more the denominator (that depends on the 2PCF) 
than the 3PCF, 
what results in a significant drop of the reduced bispectrum on 
non-linear scales.
On large-scales PT holds (see dotted line). Note that the fiducial 
model matches simulations, but
exhibits a bump on $k \approx 1-2 \impc$, that is not seen in N-body results.
\label{fig:qeqk_paper_hb}}
\end{figure}

\begin{figure}[htb]
\figurenum{6}
\epsscale{1.}
\plotone{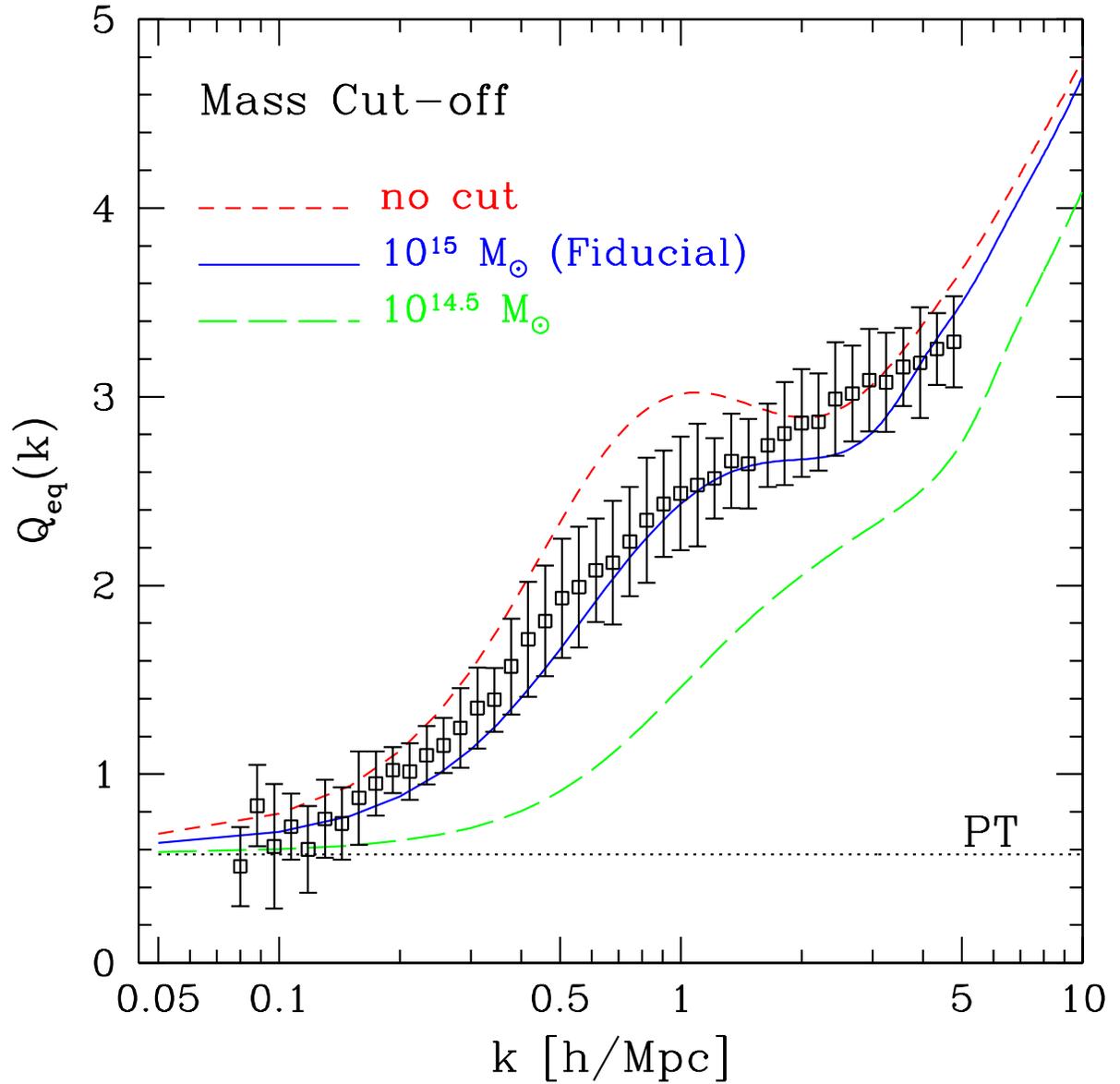}
\caption{Same as Fig.$\ref{fig:qeqk_paper_hb}$, but for the mass cut-off.
Excluding the largest mass haloes significantly lowers the amplitude of 
the reduced bispectrum.
\label{fig:qeqk_paper_mc}}
\end{figure}

\begin{figure}[htb]
\figurenum{7}
\epsscale{1.}
\plotone{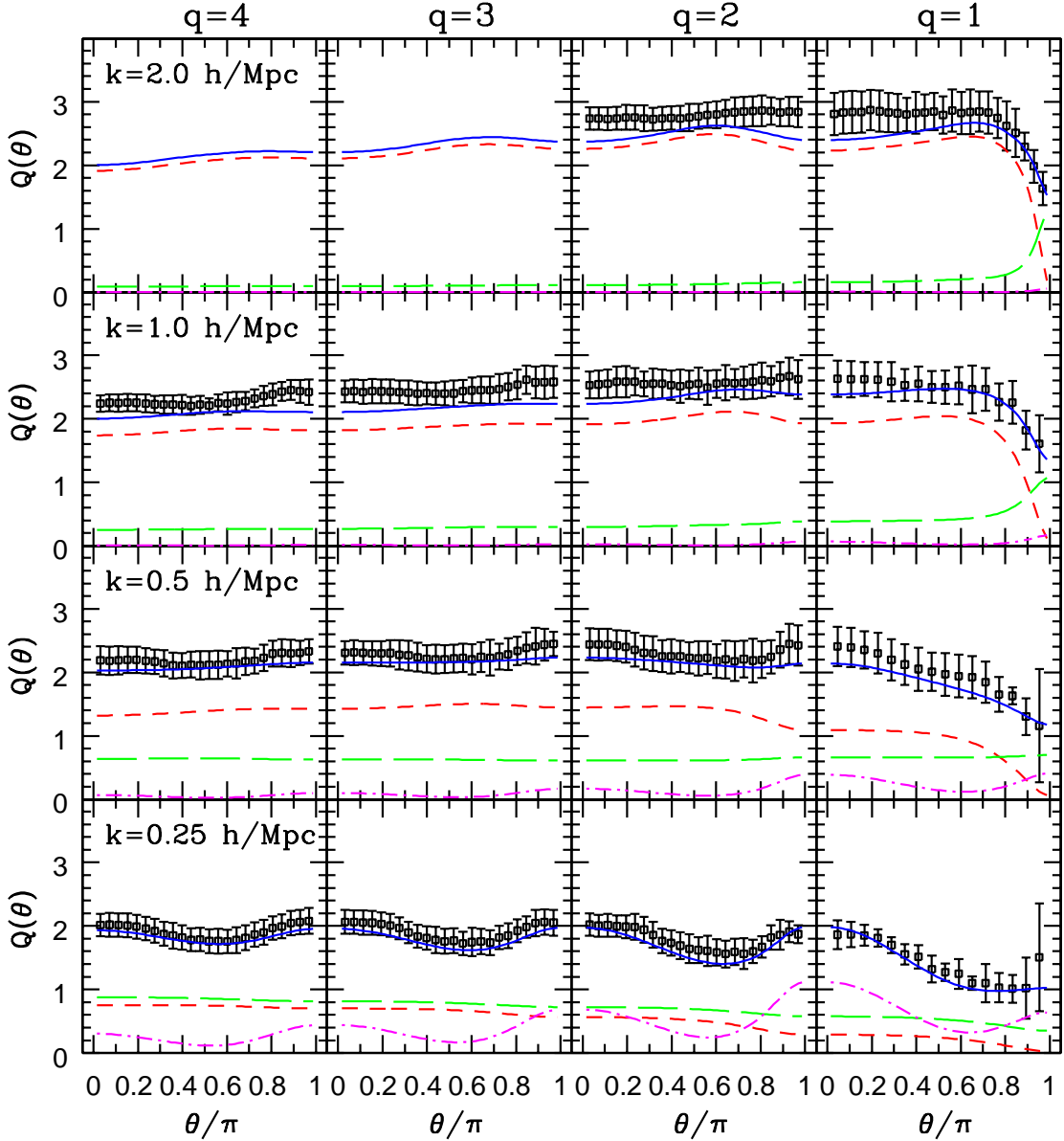}
\caption{Reduced bispectrum for different triangle configurations.
Rows show increasingly larger scales from top to bottom, while columns evolve
towards more symmetric triangles from left ($q=4$) to right ($q=1$, 
isosceles triangle),
where $q \equiv k_2/k_1$, and $k \equiv k_1$. Lines show the total 
(solid), one-halo (short-dashed),
two-halo (long-dashed), and three-halo (dot-dashed) terms. 
Note the continuity of all the
lines through plotted triangle configurations (from left to right). 
This ``boundary condition'' 
can be expressed as $Q(k,q,\theta/\pi=1) = Q(k,q-1,\theta/\pi=0)$, 
since these configurations describe the same triangle: the fact
that the lines continue as expected shows the accuracy of our
integration. It is seen that  large scales show a more 
pronounced configuration dependence as dictated by the three-halo term.
Triangles on small-scales are rather flat, except for the ``shoulder'' 
displayed by the isosceles triangles. 
\label{fig:q3k_scales}}
\end{figure}

\begin{figure}[htb]
\figurenum{8}
\epsscale{1.}
\plotone{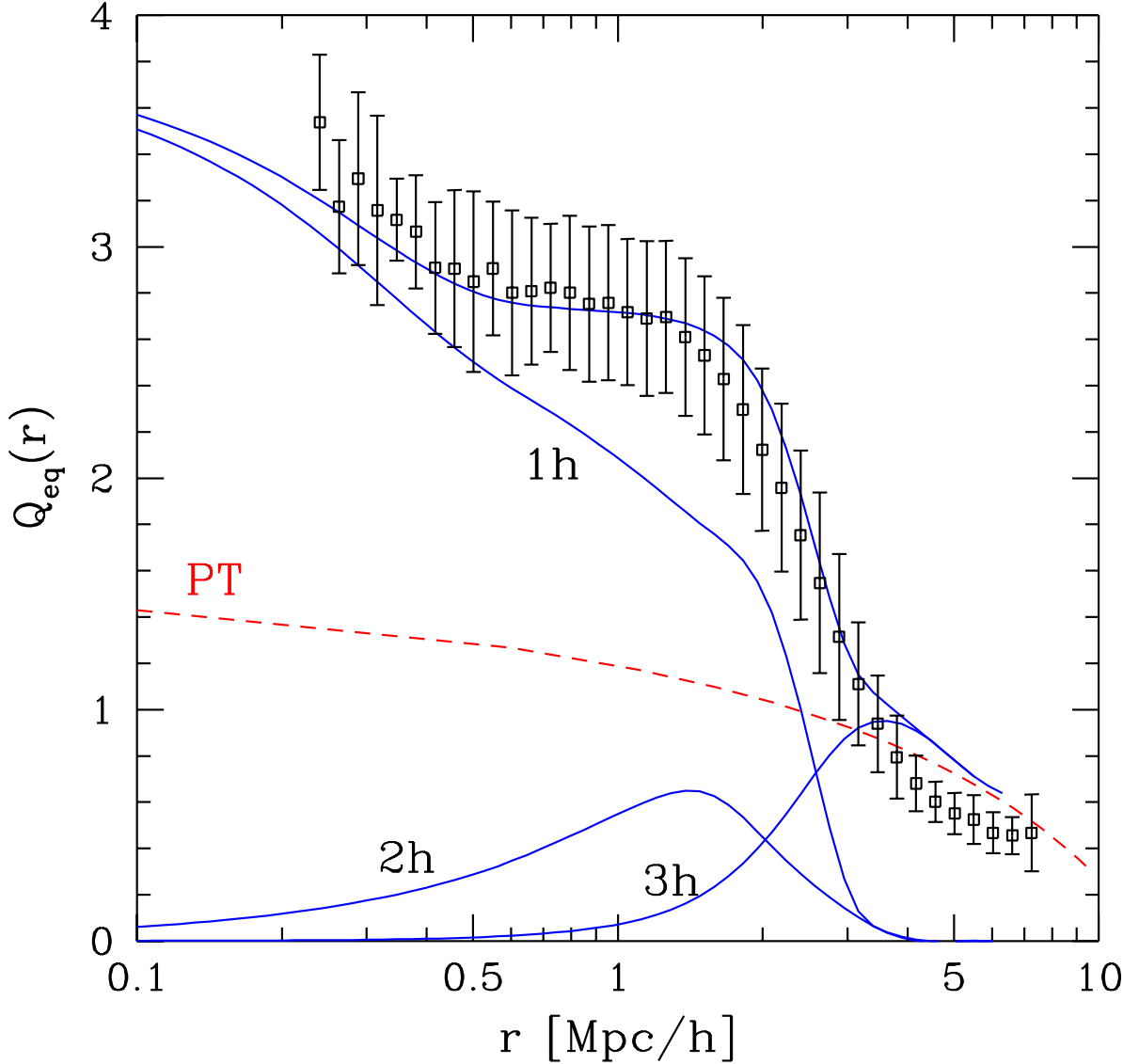}
\caption{Reduced 3PCF for equilateral triangles: the fiducial halo model 
(upper solid line) shows
good agreement with simulations. 
The small-scale enhancement is determined by the one-halo term,
while the amplitude on the transition scales $\R \approx 1-3 ~\mpc$ 
reflects the non-trivial interplay between different halo terms 
(lower solid lines). 
The good agreement found with
N-body in this range is a remarkable success of the halo model.
The observed mismatch between model and simulations on larger scales is 
due to the model slightly overestimating the 2PCF 
(that goes into the denominator of $Q_{eq}$) 
on that range of scales (see Fig.$\ref{fig:xi2_paper_hb}$).
\label{fig:qeq_paper}}
\end{figure}

\begin{figure}[htb]
\figurenum{9}
\epsscale{1.}
\plotone{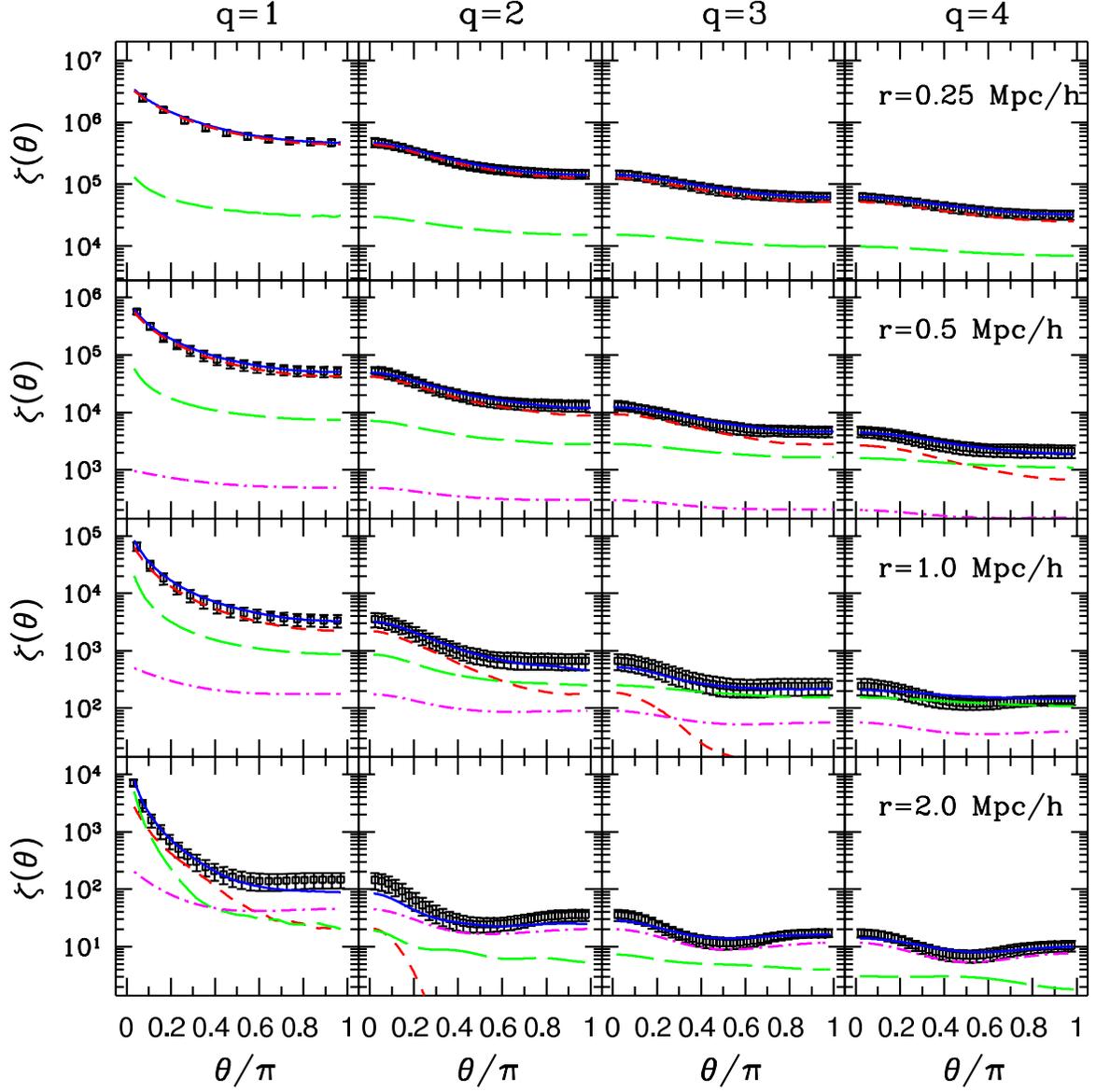}
\caption{3PCF for different triangle configurations: lines go as in 
Fig.$\ref{fig:q3k_scales}$.
Halo model predictions are shown to be in good agreement with simulations 
(symbols).
On scales $\R \simgt 1 ~\mpc$ the model shows a flatter configuration 
dependence
than simulations, but they still agree within errorbars for most of the cases. 
Note that in real space the triangle ``boundary condition'' (see text and
Fig.$\ref{fig:q3k_scales}$ for details) is expressed 
as $\zeta(r,q,\theta/\pi=1) = \zeta(r,q+1,\theta/\pi=0)$, 
where $q \equiv r_2/r_1$, and $r \equiv r_1$.
\label{fig:zeta_scales}}
\end{figure}

\begin{figure}[htb]
\figurenum{10}
\epsscale{1.}
\plotone{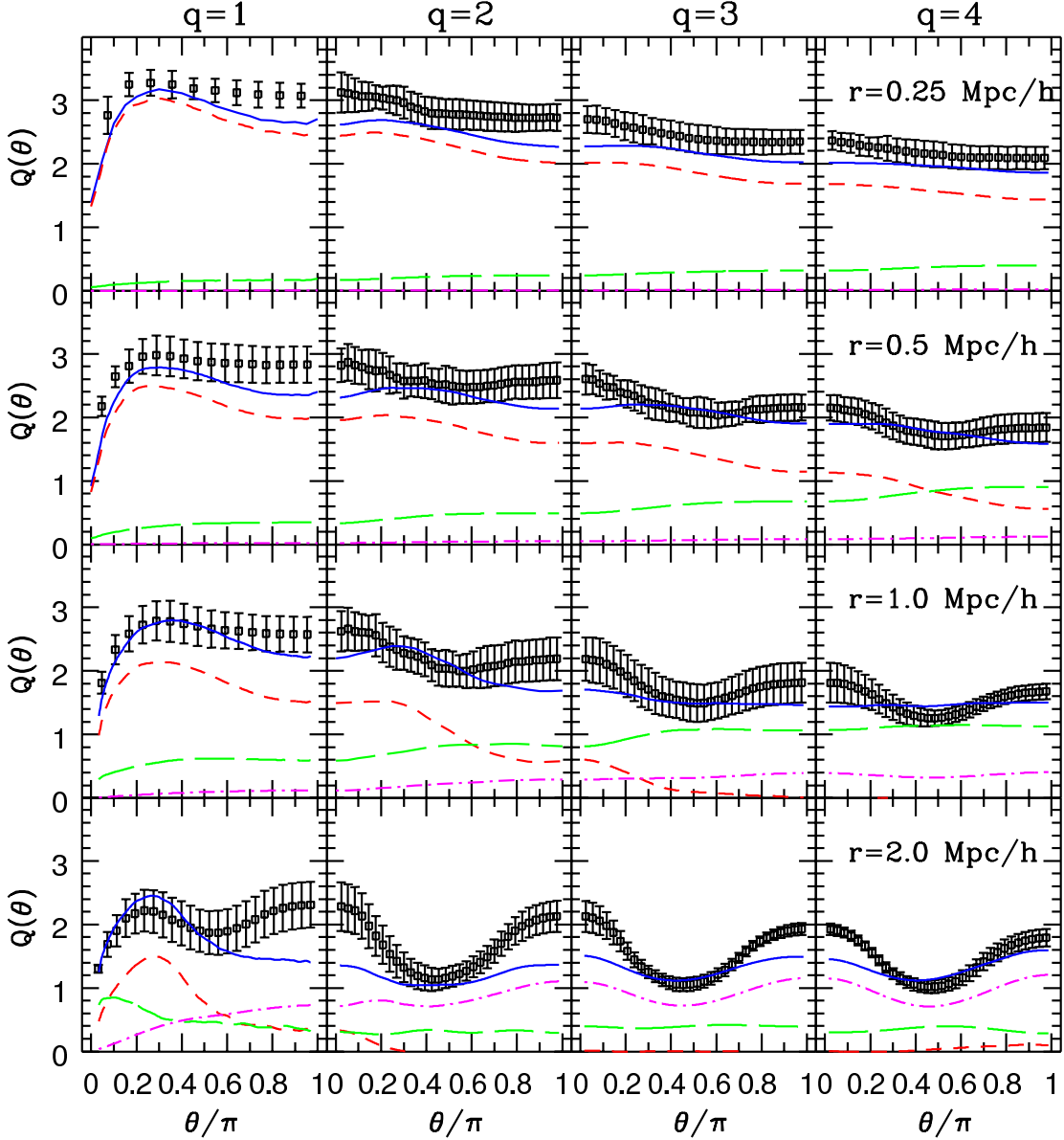}
\caption{Same as Fig.$\ref{fig:zeta_scales}$ but for the reduced 3PCF.
Halo model reproduces simulations reasonably well, specially for 
intermediate angles,
what is in agreement with Fig.$\ref{fig:qeq_paper}$ that shows equilateral 
triangles alone.
The discrepancy with simulations becomes more evident in this ratio statistic, 
as it  combines the complicated configuration dependences of both 2PCF and 
3PCF.  As in Fig.$\ref{fig:zeta_scales}$, 
the discrepancy becomes more significant for $\R \simgt 0.5~\mpc$.
\label{fig:q3_scales}}
\end{figure}

\begin{figure}[htb]
\figurenum{11}
\epsscale{1.}
\plotone{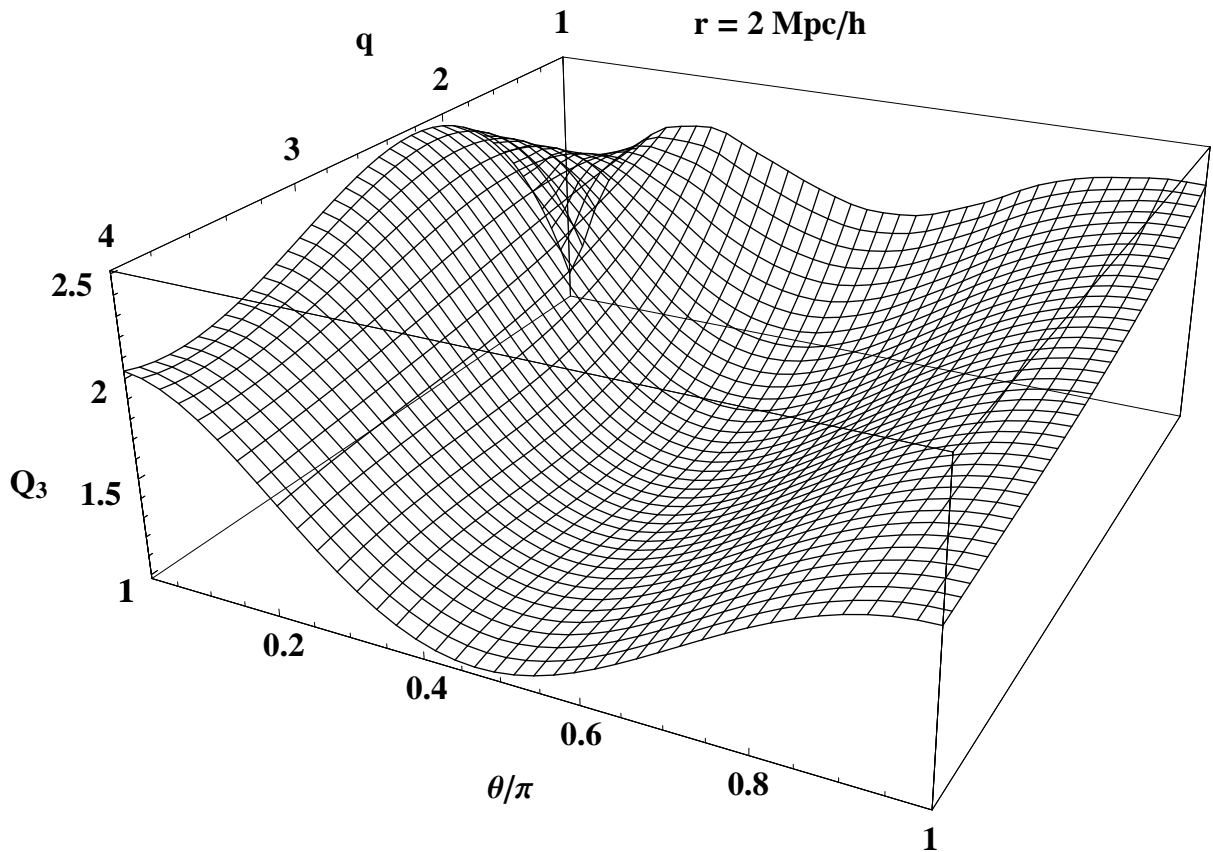}
\caption{Smoothly interpolated surface showing the 
configuration dependence of the reduced 3PCF at $\R = 2 ~\mpc$ from 
N-body simulations. 
It exhibits a symmetric convex shape, what is in qualitative 
agreement with predictions from the three-halo term in the halo model. 
For isosceles triangles this symmetry is broken at small-angles
due to non-linear effects, as reproduced by the one-halo 
contribution (see Fig.$\ref{fig:q3_scales}$).
\label{fig:q3_3d_sims_r2p0}}
\end{figure}

\begin{figure}[htb]
\figurenum{12}
\epsscale{1.}
\plotone{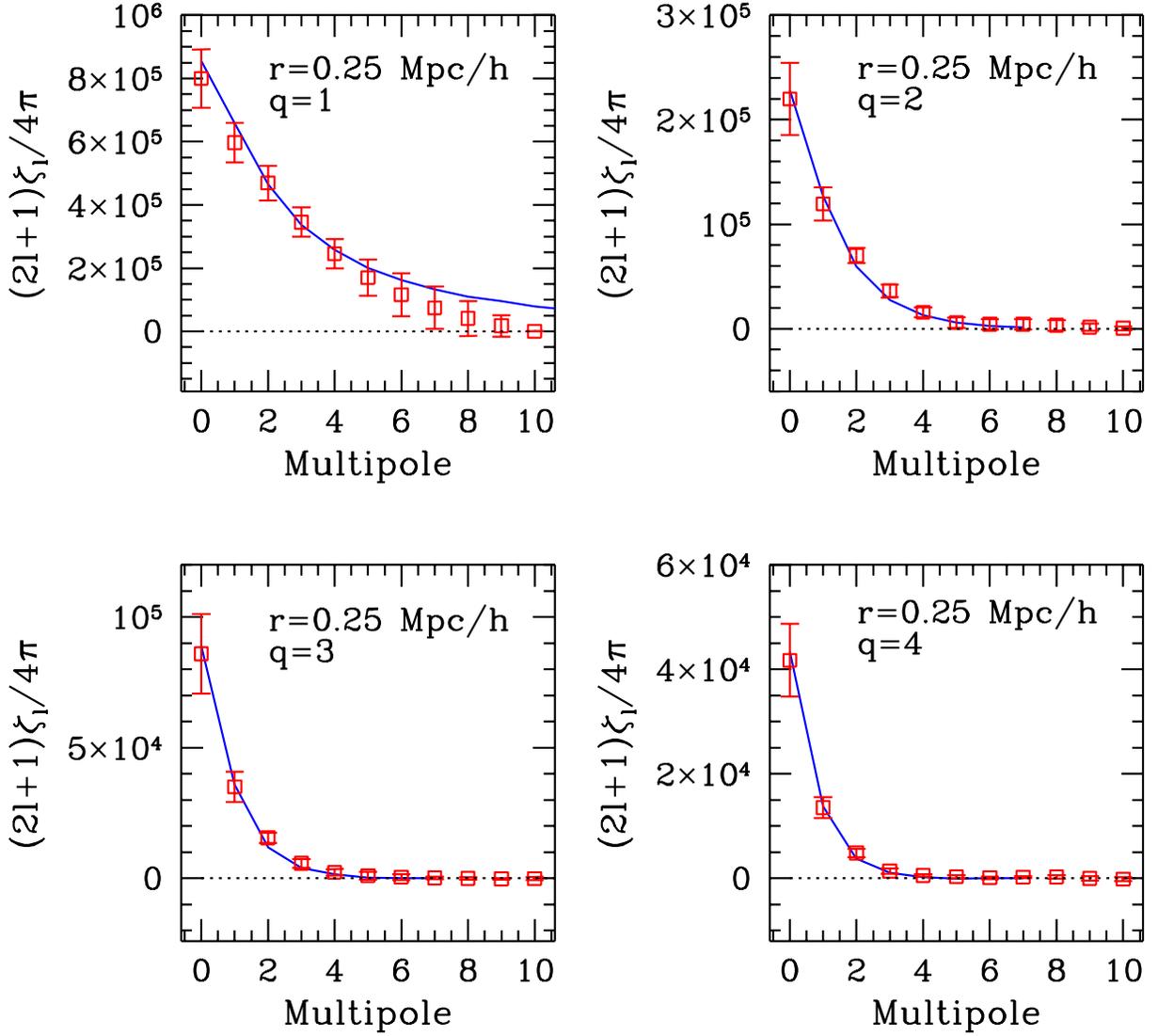}
\caption{Harmonic multipoles of the 3PCF at $\R = 0.25 ~\mpc$: 
theoretical predictions are in excellent agreement
with simulations. Only lower order multipoles ($\ell \simlt 5$, 
except for isosceles triangles) 
have a non-zero contribution. 
\label{fig:zl_0p25_comb}}
\end{figure}

\begin{figure}[htb]
\figurenum{13}
\epsscale{1.}
\plotone{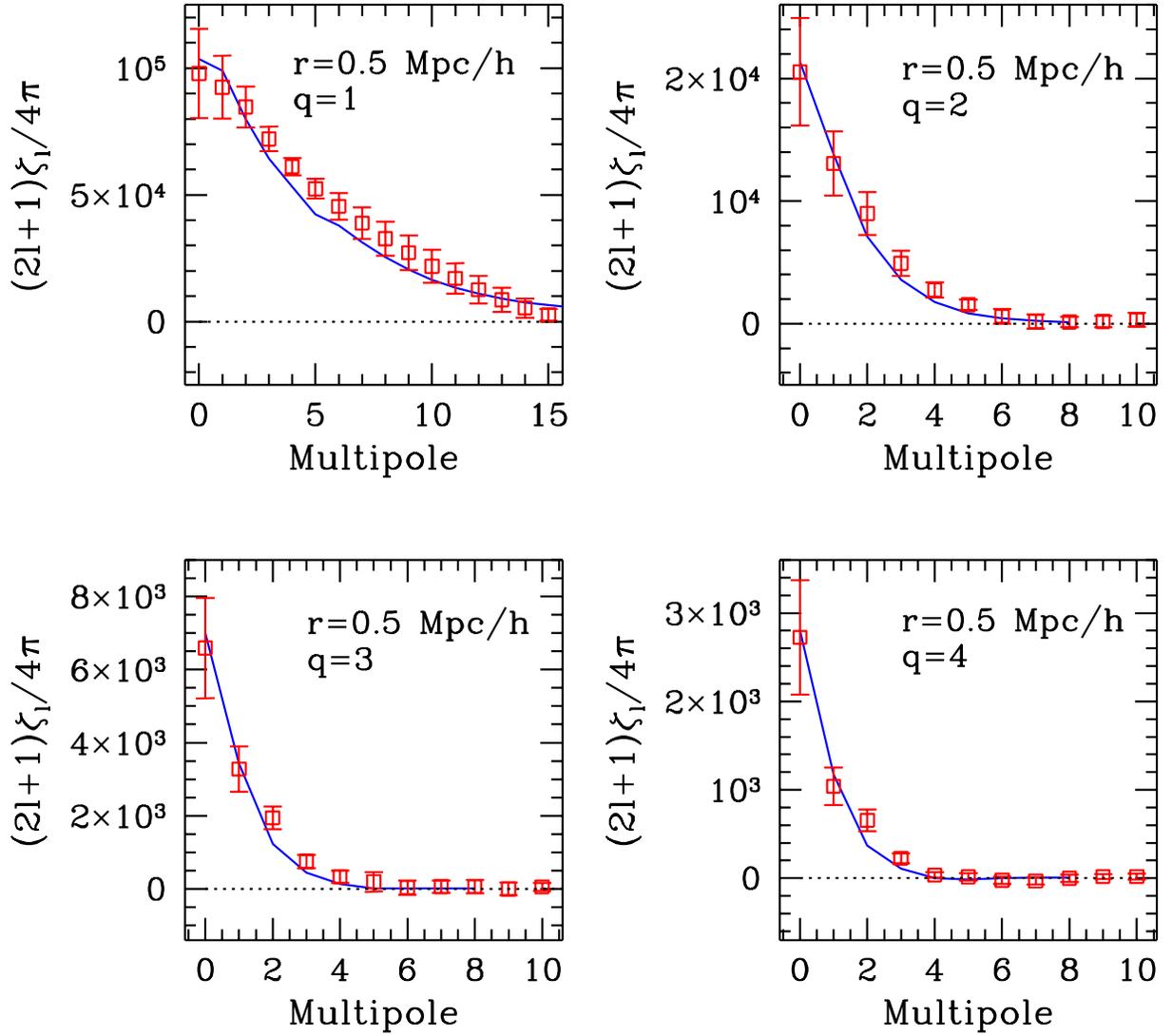}
\caption{Same as Fig.$\ref{fig:zl_0p25_comb}$ but for $\R = 0.5 ~\mpc$.
\label{fig:zl_0p5_comb}}
\end{figure}

\begin{figure}[htb]
\figurenum{14}
\epsscale{1.}
\plotone{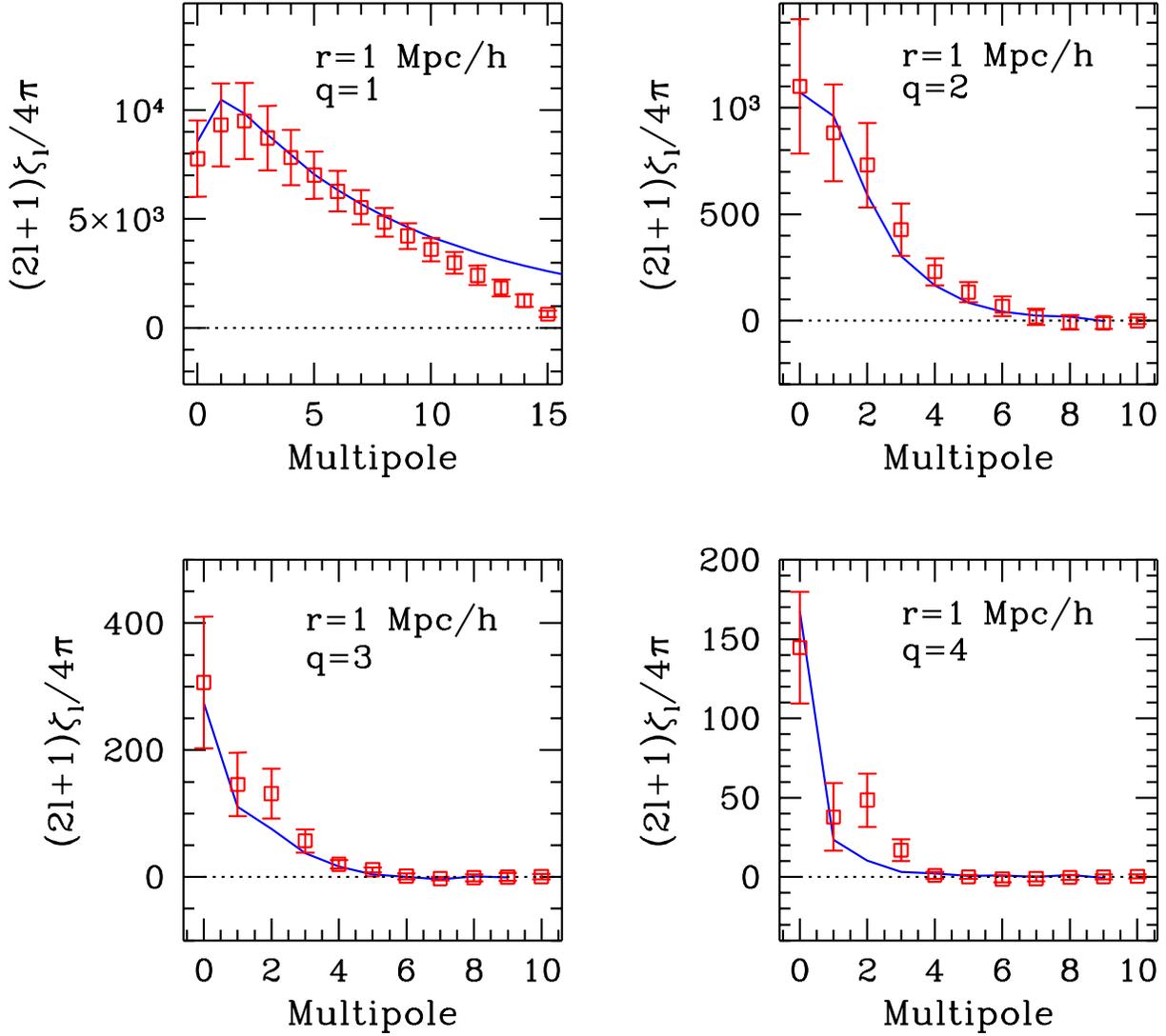}
\caption{Same as Fig.$\ref{fig:zl_0p25_comb}$ but for $\R = 1 ~\mpc$. 
Note how the higher multipoles becomes significant for
isosceles triangles ($q=1$), and
that theoretical 
predictions tend to underestimate the quadrupole moment for $q > 1$. 
\label{fig:zl_1p0_comb}}
\end{figure}

\begin{figure}[htb]
\figurenum{15}
\epsscale{1.}
\plotone{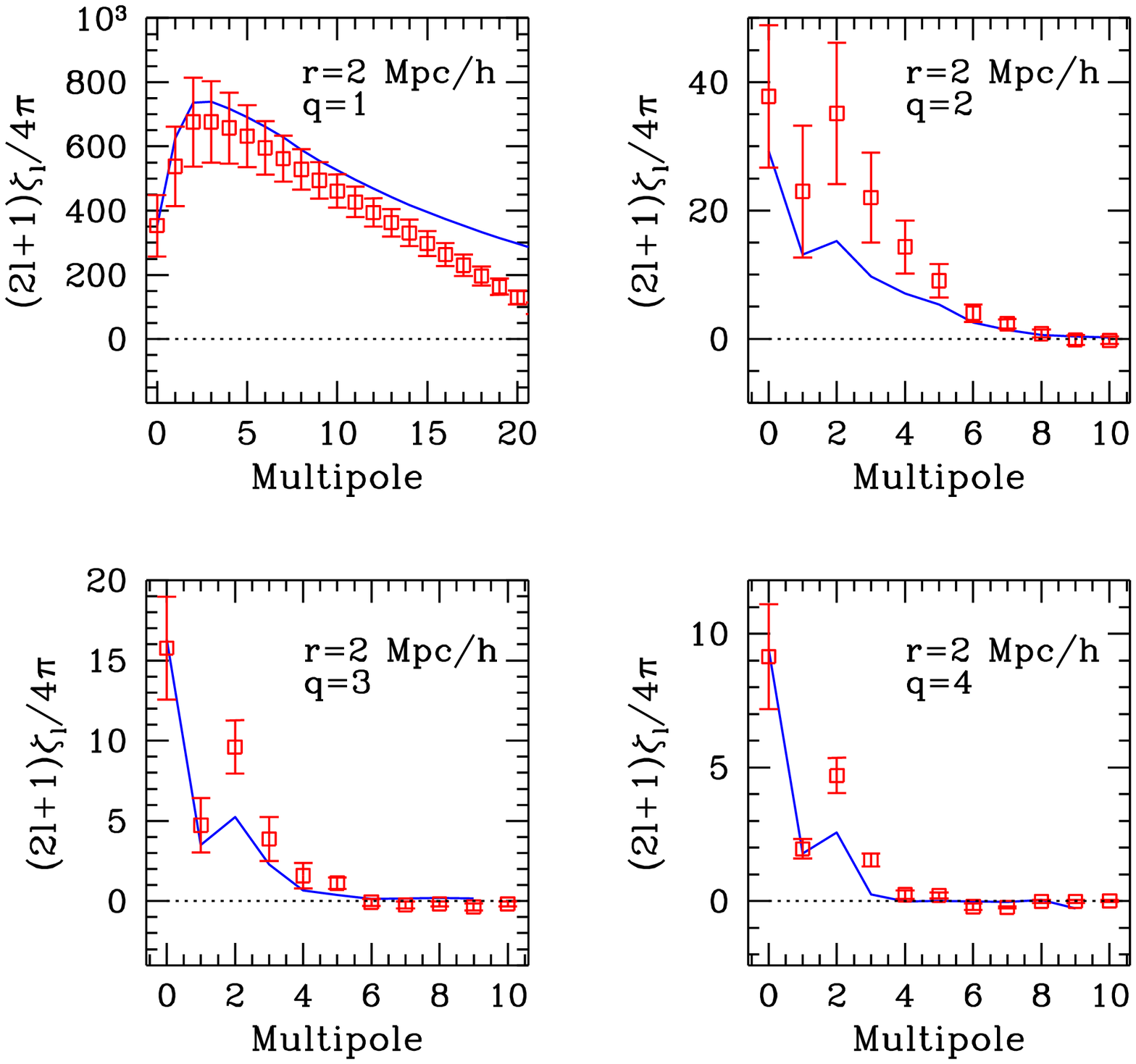}
\caption{Same as Fig.$\ref{fig:zl_0p25_comb}$ but for $\R = 2 ~\mpc$. 
In line with Fig.$\ref{fig:zl_1p0_comb}$, it shows that halo model clearly 
underpredicts
the quadrupole (and to a lesser extent, the octopole) observed in 
numerical simulations.
This {\it quadrupole deficit} explains the lack of $\theta$-dependence  
in halo model predictions for $Q(r,q,\theta)$, as shown in the lower 
rows of Fig.$\ref{fig:q3_scales}$.
\label{fig:zl_2p0_comb}}
\end{figure}

It is particularly useful to decompose the reduced 3PCF in its 
harmonic multipoles $Q_{\ell}$'s, as given by eq(\ref{eq:ql}). We have seen that
only the lowest orders have non-zero amplitude 
and thus the non-linear 3PCF can be compressed
in a few numbers (see Fig. \ref{fig:q_lterms}). 
Furthermore, assuming a model for the halo occupation 
distribution and an appropriate modeling of redshift distortions, 
one should be able to formally decompose the 3PCF of galaxies 
in the exact same way.   
In particular, observables such as the harmonic multipoles of the galaxy 
3PCF measured in large volume surveys 
should be largely uncorrelated and 
approximately Gaussian distributed, in a similar way it happens
for the angular power spectrum multipoles of the CMB anisotropy $C_{\ell}$.
 
 Thus the $Q_{\ell}$'s must be tightly related to the signal-to-noise
 eigenmodes of the 3PCF  
\citep{Scoccimarro2000, GaztanagaScoccimarro2005}. 
Moreover \cite{GaztanagaScoccimarro2005}
have shown that the highest signal-to-noise Q-eigenmodes  
measured in mock galaxy catalogs
can be efficiently used to recover bias parameters. 
It is remarkable that the first three Q-eigenmodes have
a configuration dependence that is extremely similar to
the monopole, dipole and quadrupole terms 
(compare their Fig.10 with middle panels in 
Fig.\ref{fig:q_lterms} of this paper).
This suggests that the $Q_{\ell}$'s  are indeed closely 
related to the uncorrelated modes of the galaxy 3PCF 
and thus they may simplify the procedure of constraining galaxy bias
from the 3PCF. 

Alternatively, the proposed multipole approach can be readily applied to 
other clustering statistics of the mass density field in the non-linear regime 
such as the projected mass 3PCF, that is barely affected by redshift distortions
\citep[see][for a recent implementation using Fourier series]{Zheng2004},  
or the 3PCF of the convergence field that probes the lensing potential \citep[see \eg][]{TakadaJain2003b}.

We would like to thank J.Fry, E.Gazta\~{n}aga, R.Scoccimarro and M.Takada for useful comments and discussions. 
PF ackowledges support from the spanish MEC through a Ram\'on y Cajal
fellowship and project AYA2002-00850 with EU-FEDER funding. 
This research was supported by NASA through 
ATP NASA NAG5-12101 and AISR NAG5-11996, 
as well as by NSF grants AST02-06243 and ITR 1120201-128440.
JP acknowledges support by PPARC through PPA/G/S/2000/00057.






\end{document}